# Multi-Stage Segmentation and Cascade Classification Methods for Improving Cardiac MRI Analysis

Vitalii Slobodzian[1], Pavlo Radiuk[1,*], Oleksander Barmak[1] and Iurii Krak[2,3]

[1] *Khmelnytskyi National University, 11, Institutes str., Khmelnytskyi, 29016, Ukraine*
[2] *Taras Shevchenko National University of Kyiv, 64/13, Volodymyrska str., Kyiv, 01601, Ukraine*
[3] *Glushkov Cybernetics Institute, 40, Glushkov Ave., Kyiv, 03187, Ukraine*

**Abstract**
The segmentation and classification of cardiac magnetic resonance imaging are critical for diagnosing heart conditions, yet current approaches face challenges in accuracy and generalizability. In this study, we aim to further advance the segmentation and classification of cardiac magnetic resonance images by introducing a novel deep learning-based approach. Using a multi-stage process with U-Net and ResNet models for segmentation, followed by Gaussian smoothing, the method improved segmentation accuracy, achieving a Dice coefficient of 0.974 for the left ventricle and 0.947 for the right ventricle. For classification, a cascade of deep learning classifiers was employed to distinguish heart conditions, including hypertrophic cardiomyopathy, myocardial infarction, and dilated cardiomyopathy, achieving an average accuracy of 97.2%. The proposed approach outperformed existing models, enhancing segmentation accuracy and classification precision. These advancements show promise for clinical applications, though further validation and interpretation across diverse imaging protocols is necessary.

## 1. Introduction

Cardiovascular disease (CVD) remains the primary cause of global mortality, accounting for approximately 17.9 million deaths annually [1]. Its substantial impact highlights an urgent demand for effective diagnostic tools to detect and manage heart-related pathologies early. Cardiac magnetic resonance imaging (MRI) has established itself as the "gold standard" in cardiac diagnostics, offering non-invasive, high-resolution images of heart structures and functions. These capabilities make MRI indispensable for identifying conditions such as myocardial infarction, cardiomyopathies, and structural abnormalities [2, 3].

Despite its strengths, cardiac MRI faces considerable challenges. The heart's intricate anatomy and its continuous motion due to respiration and heartbeat introduce artifacts that compromise image clarity. Additional factors, such as the presence of metal implants or equipment-induced distortions, further complicate accurate image interpretation [4, 5]. These issues often require labor-intensive image preprocessing and corrections, thereby increasing the cost and time required for analysis.

Artificial intelligence (AI) has emerged as a transformative technology in medical imaging, demonstrating its ability to automate complex tasks and identify subtle abnormalities that may elude human observers. Deep learning (DL), in particular, has shown remarkable potential for tasks such as image segmentation and classification, offering high accuracy and consistency [6]. However, the integration of AI into medical workflows faces several obstacles, including the need for extensive annotated datasets, concerns about data privacy, and challenges in adapting AI models to diverse clinical environments [7].

The primary issue in cardiac MRI processing is the difficulty in achieving accurate segmentation and classification of MRI scans due to motion artifacts, complex heart anatomy, and existing model limitations. Existing solutions often struggle with issues like image artifacts, poor segmentation in complex cases, and the inability to accurately classify various heart conditions due to segmentation errors. Thus, this study aims to address these challenges by introducing an innovative approach to cardiac MRI analysis. Specifically, the objective is to design novel methods that deliver highly accurate segmentation and classification performance, ultimately advancing clinical decision-making.

The structure of the paper is as follows: Section 2 reviews the state-of-the-art techniques in cardiac MRI segmentation and classification, highlighting advancements and limitations. In Section 3, the manuscript introduces a multi-stage segmentation process using U-Net and ResNet models, followed by a cascade classification system. Section 4 presents improved segmentation accuracy through mask localization and postprocessing, alongside high classification precision. Finally, Section 5 summarizes the study's findings, emphasizing its contributions to enhancing cardiac MRI analysis and discussing potential limitations and future research directions.

## 2. Related works

DL has completely transformed medical image analysis by uncovering complex patterns in data that traditional methods struggle to identify. Models like U-Net [8] and ResNet [9] have been instrumental in achieving accurate image segmentation, even when trained on limited datasets. U-Net's encoder-decoder architecture, for instance, efficiently captures both global and local image features. However, these models often demand significant computational resources and rely on substantial training data to achieve optimal performance [10].

Recent trends emphasize building trust in AI systems by introducing human-in-the-loop [11] and human-centric approaches [12]. While these hybrid techniques improve interpretability and reliability, they increase the complexity of deployment. Additionally, combining deep learning with traditional methods, such as active contour modeling, enhances segmentation precision but adds to computational overhead [13].

In the field of cardiac MRI, multimodal approaches that integrate data from various imaging modalities, such as CT and MRI, have shown promise [14]. While these methods improve segmentation outcomes, their reliance on datasets from different imaging sources creates significant integration challenges [15]. For instance, Hu et al. [16] developed a deeply supervised network paired with a 3D Active Shape Model that reduces manual initialization efforts. Despite its effectiveness, the method's high computational demands and lack of validation across imaging protocols limit its broader applicability. da Silva et al. [17] introduced a cascade approach utilizing DL models for automatic segmentation of cardiac structures in short-axis cine-MRIs, achieving enhanced segmentation accuracy; however, it may face limitations such as increased computational complexity and reduced generalizability due to reliance on high-quality training data.

In addition, recent enhancements to U-Net, such as attention mechanisms [18] and residual connections [19], have further boosted their performance in cardiac MRI segmentation. These improvements allow the model to better focus on relevant regions and handle variations in heart anatomy. However, challenges remain in terms of computational efficiency and robustness to imaging artifacts.

Segmentation and classification are often treated as isolated tasks, but recent works aim to combine these processes. Sander et al. [20] addressed segmentation errors with a corrective framework that requires manual intervention, increasing workflow complexity. Ammar et al. [21] designed a combined segmentation-classification pipeline for diagnosing heart diseases, but its reliance on high-quality segmentation introduces additional training burdens. Similarly, Zheng et al. [22] utilized semi-supervised learning for explainable classification but encountered issues with motion artifacts. Zhang et al. [23] leveraged dilated convolutions for multi-scale segmentation, though their method struggled with overfitting and resource-intensive training.

Existing approaches to cardiac MRI face several unresolved issues, including dependency on high-quality data, poor generalizability across diverse clinical environments, and the high computational cost of model training and deployment. These limitations hinder the practical application of DL in cardiac MRI analysis.

The goal is to enhance the accuracy of heart structure segmentation and improve the classification of conditions such as hypertrophic cardiomyopathy, myocardial infarction, and dilated cardiomyopathy. The main contributions of this research are as follows:
- A multi-stage segmentation method combining U-Net and ResNet DL models for localizing and segmenting heart structures, followed by postprocessing with Gaussian smoothing to refine contours and reduce artifacts.
- An MRI classification method based on the DL cascade model for distinguishing between heart conditions by leveraging segmented MRI data.
- Significant improvement in segmentation accuracy, achieving a Dice coefficient of up to 0.974 for left ventricle (LV) and 0.947 for right ventricle (RV) segmentation.

## 3. Methods and materials

In this study, we introduce a novel approach to the segmentation and classification of MRI scans, involving a multi-stage process, as illustrated in Figures 1.

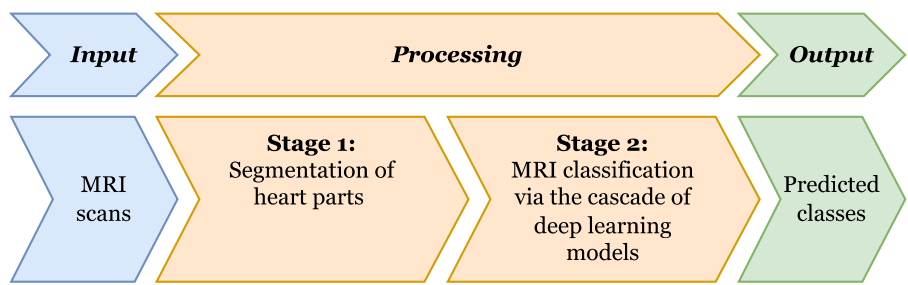

**Figure 1:** The scheme of the proposed approach: the process flow of MRI scans, starting with heart part segmentation, followed by classification using a cascade of DL models, and ending with predicted class outputs.

The proposed approach is divided into two key stages. In the first stage, relevant heart parts are segmented to extract critical anatomical features. In the second stage, a cascade of DL models [24] is employed to classify the MRI scans, ultimately producing the predicted classes. The following subsections detail each stage of the process, along with the materials and techniques used.

The first stage of the process is presented as a novel method of MRI segmentation, while the second stage is formalized as a new method of MRI classification. Below, we describe all stages of the proposed approach in detail.

### 3.1. Method of MRI segmentation

The proposed method for heart segmentation on MRIs involves three key steps: localization, mask generation, and post-processing to refine contours. First, existing masks are split into binary masks for the myocardium, LV, and RV with a DL model used to identify the region for each fragment. Then, DL helps refine the contours, and finally, the masks are combined into a single mask and resized to their original dimensions for improved accuracy.

These steps together provide an integrated approach (Figure 2), which increases the accuracy of heart segmentation on MRI scans.

Below is a detailed description of each step of the method.

The *input data* for the process in the image consists of MRI scans of the heart, where masks representing different heart structures are provided. These masks depict the LV, RV, and myocardium as distinct areas for analysis.

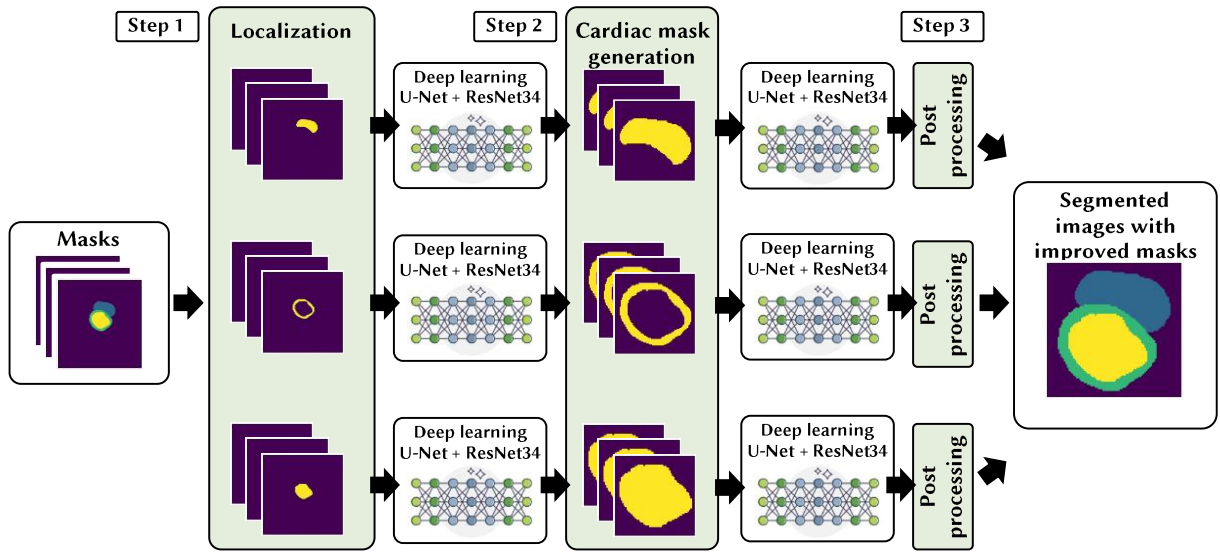

**Figure 2:** Scheme that demonstrates the proposed method of segmenting heart structures from MRI scans. Masks for the heart's LV, RV, and myocardium are localized using the U-Net and ResNet DL models. These localized masks are further refined through cardiac mask generation, followed by postprocessing to improve accuracy and produce segmented images with improved masks.

*Step 1.* The localization part consists of decomposing the existing masks into separate binary masks for different heart structures: myocardium, LV, and RV (Figure 3).

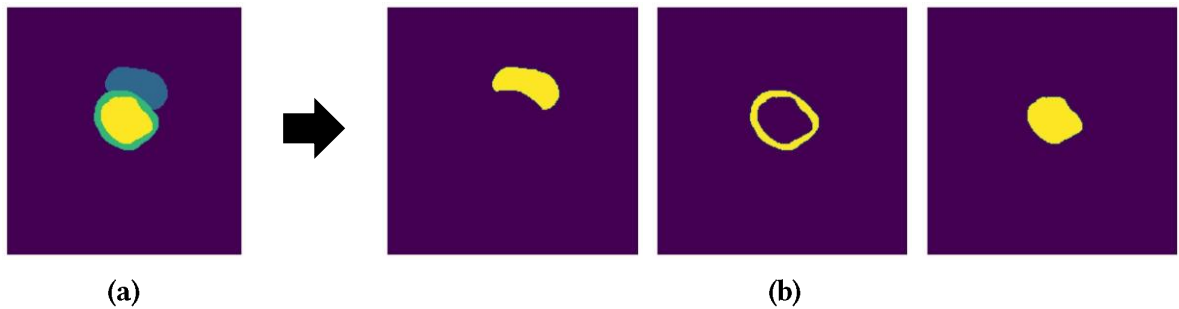

(a)          (b)

**Figure 3:** Decomposition of a general mask (a) into three binary masks (b).

This process allows each heart structure to be processed separately, improving segmentation accuracy. Each binary mask focuses on a specific heart structure, where relevant pixels are marked as 1, and all others are 0. This separation helps DL models target individual structures, reducing interference from other parts of the image and simplifying the segmentation task, which boosts accuracy and reduces computational complexity.

For each mask, a separate DL model is trained to detect the location of a specific heart fragment, working like an object detector to identify boundaries within the MRI scan. For example, the model trained for the LV focuses only on locating that specific structure.

The models are trained using the Fastai library [25] and pre-trained networks built on U-Net [8] and ResNet [9] architectures, with the ResNet-34 version (34 layers) being used in this study. Images are resized for uniformity before training, and the model is trained for 10 epochs, followed by fine-tuning and an additional 10 epochs. This method improves accuracy by adjusting parameters, and the resulting masks help center and localize the heart structures by adjusting the image's aspect ratio and adding a 15% frame for better focus. The localization result for the LV is shown in Figure 4.

As an outcome, the first phase yields localized images with marked regions of interest: the myocardium, LV, and RV.

*Step 2.* For cardiac mask generation, there were three separately trained models for each heart structure. These models take the localized images from *step 1* as input and perform detailed region delineation of each heart structure.

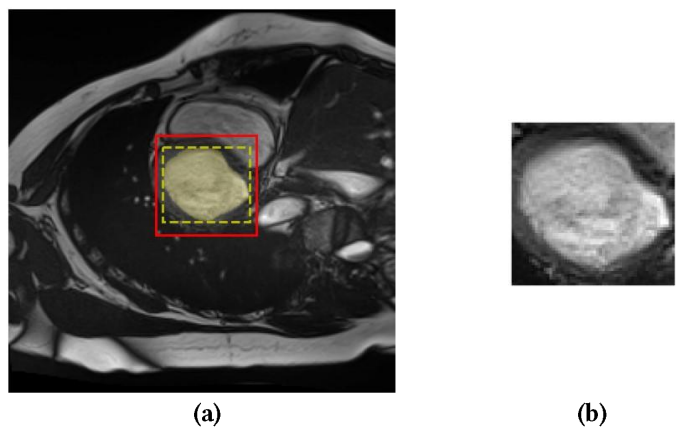

(a) (b)

**Figure 4:** An example of the localization result: (a) yellow mask – LV, yellow dashed frame – LV area, red frame – final localization area; (b) localized image of the LV area.

Training here follows the same approaches and technologies as in *step 1*. Image localization helps to operate with less data, boosting accuracy in determining heart structure contours. This localization helps avoid noise and unrelated structures, allowing the DL model to capture finer details, which is essential for this step's accuracy. Figure 5 shows original input image, samples of input localized images, and output masks from *step 2*.

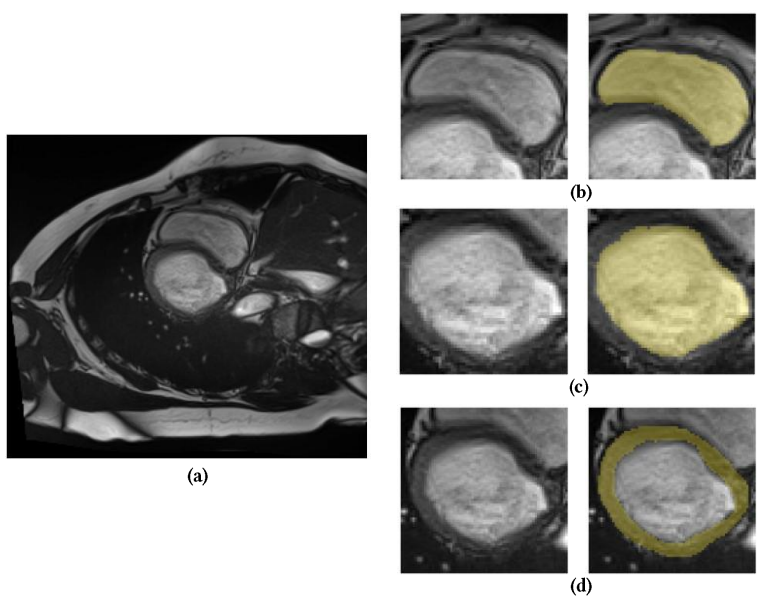

**Figure 5**: Segmentation results: (a) input image, (b) RV mask, (c) LV mask, and (d) myocardium mask.

Therefore, the output of *step 2* is segmented images containing masks of separately defined areas: the myocardium, LV and RV.

*Step 3.* Postprocessing focuses on refining and improving the quality of the generated masks. Since the models are trained on uniformly resized images, they must be scaled back to their original dimensions for proper comparison with the ground truth masks. However, simple resizing can cause detail loss and artifacts, which affects the final evaluation. To address this, smoothing methods that create smooth pixel transitions for a more natural appearance when resizing are used. In our case,

Gaussian smoothing offered an acceptable balance between performance and efficiency. It is formalized by the following formula:

$$G(x,y) = \frac{1}{2\pi\sigma^2} e^{-\frac{x^2+y^2}{2\sigma^2}}, \quad (1)$$

where $G(x,y)$ is the Gaussian filter value in point $(x,y)$, $\sigma$ stands for standard deviation, which specifies the intensity of smoothing, $(x,y)$ represent pixel coordinates.

Linear regression is utilized to identify the optimal value automatically of $\sigma$ in formula (1) for each image size.

Finally, the *output* data of the proposed method are segmented images with improved masks in their original size for more correct comparison with expert masks.

### 3.2. Method of MRI classification

The proposed classification method detects abnormalities in LV and RV or confirms a normal state by analyzing MRI scans across different cardiac cycle stages. Structured in multiple levels to minimize class confusion and improve generalization, it incorporates critical anatomical features such as tissue density, ventricular volume, and dynamic myocardium thickness.

By leveraging segmentation results from prior steps of method of MRI segmentation and combining MRI scans and segmentation masks from both diastolic and systolic phases, the model captures both geometric and texture details essential for accurate diagnosis. Each heart segment is represented in separate RGB channels, aiding the DL model in analyzing structural and tissue heterogeneity, with images interpolated to a consistent size to reduce noise and irrelevant details before classification.

Figure 6 shows the set of images that are typically fed into the DL model.

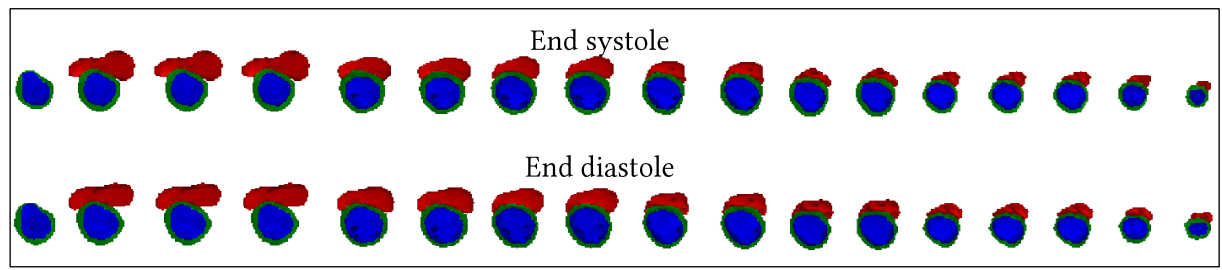

**Figure 6**: Visualizations of input data. The top row represents images from the systolic phase, while the bottom row shows images from the diastolic phase. The columns correspond to slices along the short axis. Red marks indicate the segments of the RV, blue marks – the LV, and green marks – the myocardium segment.

To address the common issue of class imbalance in medical datasets, the proposed method uses a cascading classification model, following the scheme in [24]. This approach helps improve generalization in small datasets by training binary classifiers that focus on two specific classes at a time, enhancing classification accuracy.

The cascade consists of four classifiers:
1. The *first* classifier separates LV pathologies from RV pathologies and normal conditions, allowing the model to focus on general LV features.
2. The *second* classifier distinguishes between RV abnormalities and normal conditions, further refining the model's accuracy.
3. The *third* classifier differentiates hypertrophic cardiomyopathy from other LV pathologies.
4. The *fourth* classifier separates myocardial infarction-related pathologies from dilated cardiomyopathy, which are often hard to tell apart, enabling the model to better distinguish between them.

Figure 7 illustrates the application of all four classifiers for pathology identification.

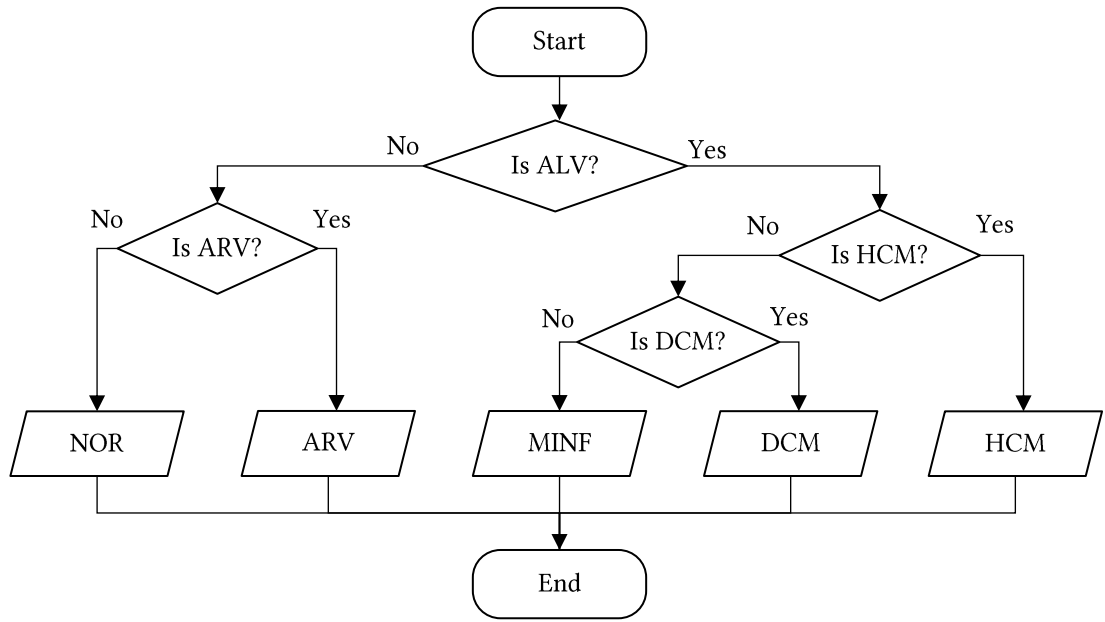

**Figure 7**: Algorithm for cascading application of binary classifiers.

The proposed classifiers utilize the CNN model [26] adapted for the task of binary classification. The architecture is schematically represented in Figure 8.

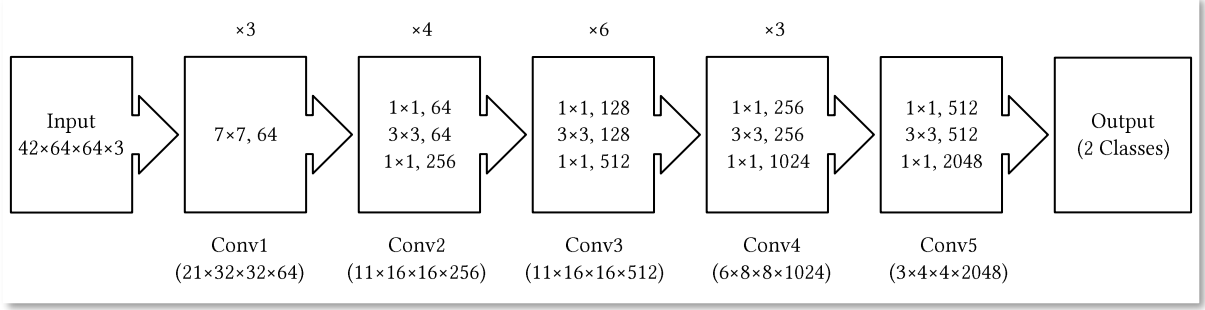

**Figure 8**: Architecture of the DL model within the proposed method for classifiers.

The model architecture has 50 layers and includes essential components like an initial convolutional layer for extracting basic features and normalization and activation layers to stabilize learning. The first layer, Conv1, uses large filters to capture basic features like edges and textures, followed by Conv2 through Conv5, which apply various filters to learn more complex and abstract details at each stage.

After these convolutional operations, global average pooling gathers all learned features into a single vector, which is then passed to the final layer responsible for binary classification. This multi-layered processing allows the model to accurately analyze both simple and complex patterns in the input data, making it highly suitable for classification tasks.

The overall method scheme is depicted in Figure 9.

The method involves the following key steps. The input data consists of modified images from the dataset, including MRI scans for each patient during both the diastolic and systolic phases.

*Step 1*: MRI scans are prepared by cropping to focus on the necessary heart segments, then resizing them to a uniform dimension. The segmentation masks and images are combined, with each heart segment placed in a separate channel.

*Step 2*: The cascade of four classifiers is trained, with each classifier trained individually. The model is compiled using the Adam optimizer and the "categorical cross-entropy" loss function, with the data split into training and validation sets. Early stopping is used during training to prevent overfitting by stopping the process if validation losses don't improve.

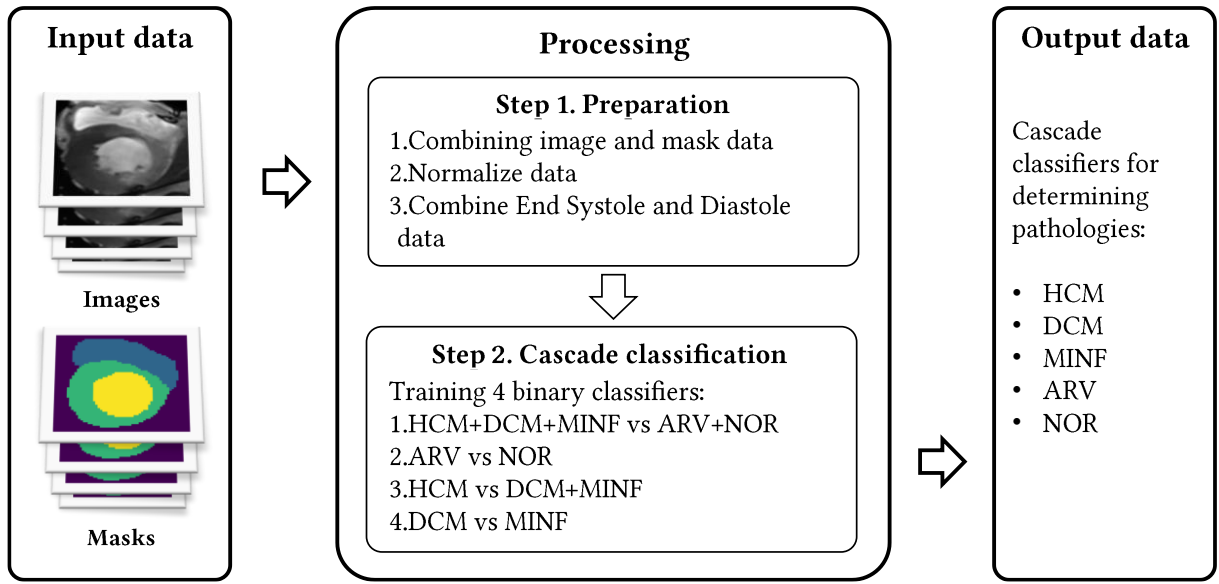

**Figure 9**: Scheme of the proposed method of classifying heart pathologies from MRI scans. The input data includes both MRI scans and masks representing different heart structures. The process involves preparing the data by combining images with masks and normalizing them, followed by a cascade classification system to identify specific heart conditions.

The output of the method is a trained cascade of classifiers that can identify the following pathologies:
1. Abnormal right ventricle (ARV).
2. Hypertrophic cardiomyopathy (HCM).
3. Previous myocardial infarction (MINF).
4. Dilated cardiomyopathy (DCM).
5. Normal state (NOR).

### 3.3. Dataset

The Automated Cardiac Diagnostic Challenge (ACDC) dataset [27] was used for both segmentation and classification tasks in this study. The dataset includes 150 patients split into five groups: healthy, myocardial infarction, dilated cardiomyopathy, hypertrophic cardiomyopathy, and right ventricular anomaly. Each patient's data includes physical parameters, images, and expert-annotated heart structure masks. While previous work [28] filtered the dataset for improved results, this study uses the original dataset. The pre-formed training and testing sets were used to ensure comparability with other studies.

### 3.4. Evaluation criteria

Experiments were conducted to evaluate each stage of the method, with models trained using consistent epochs, architecture, and data. The results were averaged over 10 training and testing cycles to ensure objectivity. Segmentation quality was measured using the Dice coefficient, which compares the overlap between predicted and expert masks. The Dice coefficient formula is as follows:

$$Dice = \frac{2 \times |A \cap B|}{|A| + |B|}, \qquad (2)$$

where $A$ is a set of pixels, $B$ is a set of pixels of true segmentation, $|A|$ represents set $A$ count, $|B|$ stands for set $B$ count, $|A \cap B|$ represents count of overlapped elements for the set $A$ and set $B$; a value of 0 in formula (2) indicates no overlap, and 1 indicates perfect alignment between the masks.

For classification accuracy, the average is calculated by considering each classifier's accuracy at every step and taking the arithmetic mean of all class accuracies to get the overall model accuracy. This approach ensures a fair comparison with other methods. The following formalizations are used for these calculations:

$$A_{\text{NOR,ARV}} = \frac{A_{Classifier\ 1} + A_{Classifier\ 2}}{2}, \tag{3}$$

$$A_{\text{HCM}} = \frac{A_{Classifier\ 1} + A_{Classifier\ 3}}{2}, \tag{4}$$

$$A_{\text{MINF,DCM}} = \frac{A_{Classifier\ 1} + A_{Classifier\ 3} + A_{Classifier\ 4}}{3}, \tag{5}$$

$$A = \frac{A_{NOR} + A_{ARV} + A_{HCM} + A_{MINF} + A_{DCM}}{5}, \tag{6}$$

where $A_{Classifier\ 1-4}$ represents the accuracy of each classifier, $A_{\text{NOR,ARV,HCM,MINF,DCM}}$ represents the classification accuracy of each class, with A being the overall accuracy of the method.

## 4. Results and discussion

### 4.1. Results for method of segmentation

The experimental results obtained to determine the accuracy of the localization, decomposition, and postprocessing stages are shown in Table 1.

**Table 1**
Computational results, i.e., values of Dice coefficient, to test the accuracy of the localization (L), decomposition (D), and postprocessing (PP) steps within the proposed segmentation method. Myo. of LV stands for the myocardium of LV. Numbers in **bold** represent higher values.

| Experiment | End diastole | | | End systole | | |
|---|---|---|---|---|---|---|
| | LV | RV | Myo. of LV | LV | RV | Myo. of LV |
| Original Images | 0.911 | 0.842 | 0.812 | 0.890 | 0.871 | 0.832 |
| L | 0.920 | 0.902 | 0.875 | 0.894 | 0.891 | 0.884 |
| D | 0.919 | 0.892 | 0.855 | 0.887 | 0.873 | 0.885 |
| L + D | 0.956 | 0.939 | 0.866 | 0.930 | 0.905 | 0.898 |
| L + D+ PP | **0.974** | **0.947** | **0.896** | **0.940** | **0.915** | **0.920** |

Moreover, the results obtained are compared with other methods (Table 2).

**Table 2**
Comparison of segmentation results with state of the art by Dice coefficient. Numbers in **bold** represent higher values.

| Approaches | End diastole | | | End systole | | |
|---|---|---|---|---|---|---|
| | LV | RV | Myocardium of LV | LV | RV | Myocardium of LV |
| Ours | **0.974** | **0.947** | 0.896 | **0.940** | **0.915** | **0.920** |
| Hu et al. [16] | 0.968 | 0.946 | **0.902** | 0.931 | 0.899 | 0.919 |
| da Silva et al. [17] | 0.963 | 0.932 | 0.892 | 0.911 | 0.883 | 0.901 |
| Sander et al. [20] | 0.959 | 0.929 | 0.875 | 0.921 | 0.885 | 0.895 |
| Ammar et al. [21] | 0.964 | 0.935 | 0.889 | 0.917 | 0.879 | 0.898 |

*Segmentation of original images*. In the first stage of the experiments, a model was trained to segment full MRI scans without any prior localization or decomposition. The model was trained to detect the contours of the myocardium, as well as the LV and RV, across the entire image. The results of this experiment are shown in Figure 10.

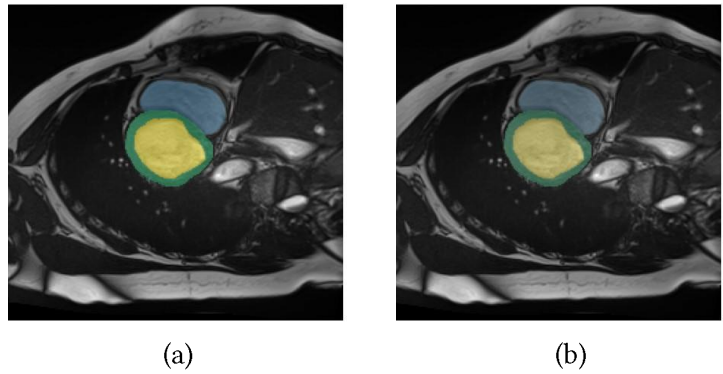

(a)            (b)

**Figure 10**: Comparison of masks for original images: (a) expert mask and (b) DL output mask.

*Localization and segmentation of original images.* The second stage of the experiments involved localization and segmentation of the original MRI scans. First models were used to determine the heart area location (with myocardium, RV, and LV). After that, the localized area was passed to the input of the DL model for detailed segmentation. An example of the result of the described experiment is shown in Figure 11.

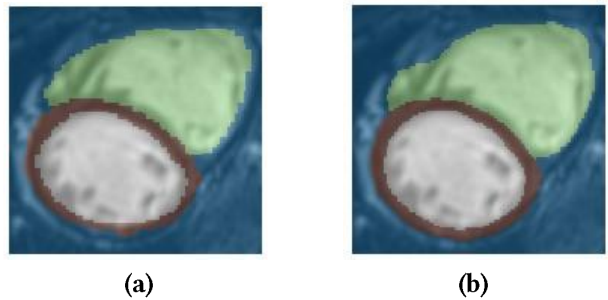

(a)            (b)

**Figure 11**: Comparison of masks for localized images: (a) expert mask and (b) DL output mask.

*Segmentation of original decomposed images.* The third stage of the experiments involved the segmentation of the original images (decomposed). The original MRI scans were divided into separate binary masks for the myocardium, LV, and RV. Separate DL models were applied for each mask, trained to identify the corresponding structures. This allowed for testing if segmentation performance increases by splitting the task into separate parts without employing preliminary localization. An example of the result of this experiment is shown in Figure 12.

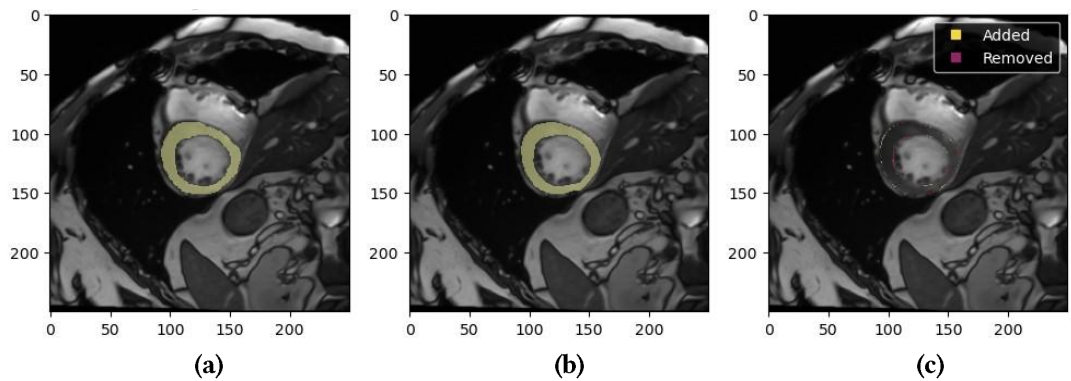

(a)            (b)            (c)

**Figure 12**: Comparison of masks for images in the original size with mask decomposition: (a) expert mask, (b) DL output mask, and (c) difference between masks.

*Localization and segmentation of decomposed images.* The fourth stage of the experiments involved localization and segmentation of the decomposed images. First, for each of the binary masks (myocardium, LV, and RV), localization models were used to define the regions of these structures. The localized regions were then passed to DL models for detailed segmentation. This approach allowed us to assess the impact of preliminary localization and decomposition on segmentation accuracy. An example of the result of the described experiment is shown in Figure 13.

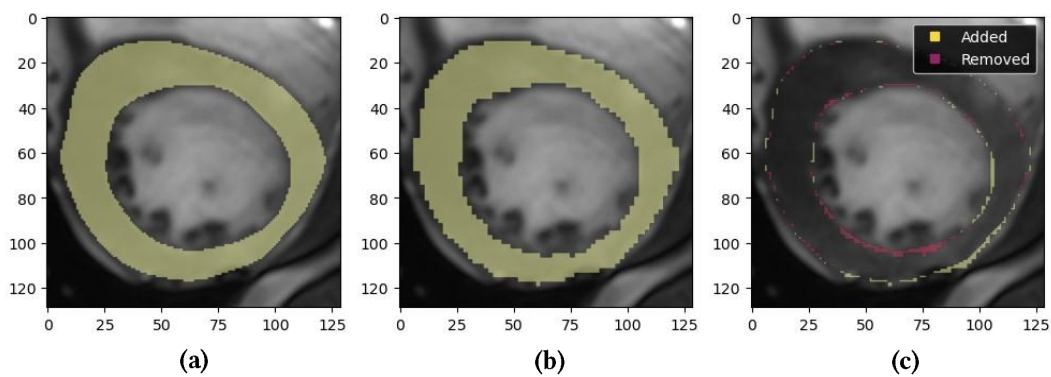

(a)                 (b)                 (c)

**Figure 13**: Comparison of masks for localized images with mask decomposition: (a) expert mask, (b) DL output mask, and (c) difference between masks.

*Localization and segmentation of decomposed images with postprocessing (proposed approach)*. At the fifth and final stage of the experiments, the decomposed images were localized and segmented, followed by postprocessing. After completing localization and segmentation for each of the binary masks, the results were processed using postprocessing to smooth transitions and reduce artifacts. The masks were returned to their original size using blurring techniques to ensure a correct comparison with the expert masks. The results are shown in Figure 14.

Therefore, the experiments have demonstrated enhanced accuracy of the proposed method, which includes localization, decomposition, and postprocessing of images. This approach provides high accuracy of segmentation of heart structures in MRI scans, which is critical for further clinical analysis and diagnosis.

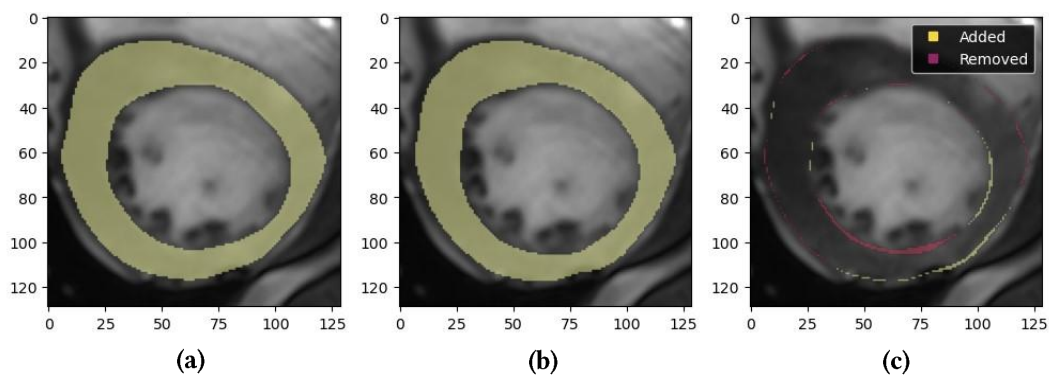

(a)                 (b)                 (c)

**Figure 14**: Comparison of masks for localized images with mask decomposition with contour enhancement: (a) expert mask, (b) DL output mask, and (c) difference between masks.

### 4.2. Results for method of classification

The proposed classification method was evaluated using several metrics, including precision, recall, $F_1$-score, and overall accuracy. For each of the four classification steps, metrics (2)–(6) were used to assess the detection and separation of various heart pathologies. Figure 15 presents the confusion matrix for each classification step, demonstrating the rate of correct, false positive, and false negative classifications.

    **Classifier 1**               **Classifier 2**               **Classifier 3**               **Classifier 4**

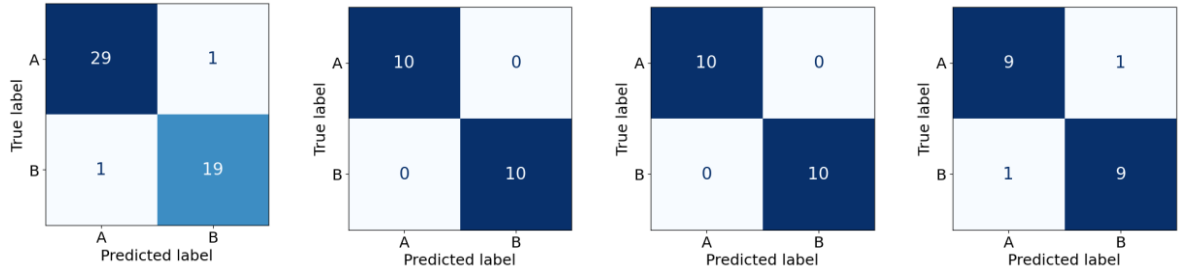

**Figure 15**: Confusion matrices for classification steps: step 1 – classifier 1, step 2 – classifier 2, step 3 – classifier 3, and step 4 – classifier 4.

Table 3 shows classification results of the proposed model at each step.

**Table 3**
Classification evaluation metrics for classifiers 1–4 obtained on steps 1–4, respectively, within the proposed classification method.

| Classifier | Classes | Precision | Recall | $F_1$-score | Accuracy |
|---|---|---|---|---|---|
| Classifier 1 | NOR+ARV | 0.95 | 0.95 | 0.95 | 0.96 |
| | MINF+HCM+DCM | 0.97 | 0.97 | 0.97 | |
| Classifier 2 | NOR | 1.00 | 1.00 | 1.00 | 1.00 |
| | ARV | 1.00 | 1.00 | 1.00 | |
| Classifier 3 | HCM | 1.00 | 1.00 | 1.00 | 1.00 |
| | MINF+DCM | 1.00 | 1.00 | 1.00 | |
| Classifier 4 | MINF | 0.90 | 0.90 | 0.90 | 0.90 |
| | DCM | 0.90 | 0.90 | 0.90 | |

The first step showed a high accuracy of 0.96 in separating LV pathologies from other cases, while the second step achieved a perfect accuracy of 1.0 for distinguishing between the normal state and RV abnormalities. The third step also achieved a perfect accuracy of 1.0 in classifying hypertrophic cardiomyopathy from other LV pathologies. Finally, the fourth step, which differentiates between previous myocardial infarction and dilated cardiomyopathy, showed an accuracy of 0.90.

Figure 16 presents the Receiver Operating Characteristic (ROC) curves for each of the four classification steps, illustrating the relationship between sensitivity (True Positive Rate) and specificity (False Positive Rate).

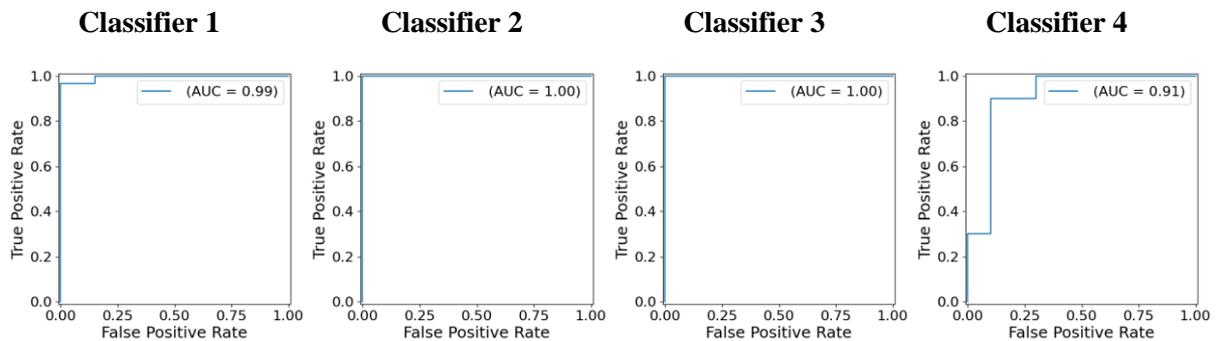

**Figure 16**: AUC curves for classification steps: step 1 – classifier 1, step 2 – classifier 2, step 3 – classifier 3, and step 4 – classifier 4.

The results obtained indicate that the proposed multi-stage segmentation and cascade classification approach delivers competitive performance in cardiac MRI analysis. The AUC values for the classification steps are consistently high, with Classifiers 1, 2, and 3 achieving near-perfect scores (0.99,

1.00, and 1.00, respectively), reflecting the model's strong ability to distinguish between classes. Classifier 4, while slightly lower with an AUC of 0.91, still demonstrates adequate performance, though there may be room for further refinement to improve classification of more challenging cases. This high classification accuracy underscores the model's effectiveness in handling various heart conditions with minimal misclassification.

A comparison of the overall accuracy of this method with the results from other authors' work is presented in Table 4.

**Table 4**
Comparison of classification results with state of the art by accuracy. Numbers in **bold** represent higher values.

| Method | Accuracy |
| --- | --- |
| Ours | 0.972 |
| Ammar et. al. [21] | 0.923 |
| Zheng et. al. [22] | 0.941 |
| Mahendra et. al. [23] | **0.998** |

Comparative analysis (Table 4) shows that our method achieves an overall classification accuracy of 0.972, positioning it closely with other state-of-the-art techniques. Although slightly lower than the highest reported accuracy of 0.998 by Mahendra et al. [23], our approach maintains a strong balance between accuracy and practical applicability, achieving improvements over several other benchmarks, including Zheng et al. [22] and Ammar et al. [21]. These results suggest that the proposed methods are robust and reliable, making them suitable for clinical applications.

### 4.3. Limitations of the proposed methods

While the proposed methods for myocardium segmentation in LV and RV show promise, there are some inherent limitations that need to be addressed. First, the model's performance can degrade significantly when processing low-quality MRI images. This is particularly noticeable when parts of the myocardium or ventricles are not fully visible, leading the model to either generate incorrect segmentations or miss the regions altogether. The model relies on detecting differences between the target structures and surrounding tissues, so poor visualization can severely affect its accuracy

Another challenge arises when the brightness levels in the images are either too low or too high. In such cases, the model might struggle to correctly identify the boundaries of the heart structures, resulting in poorly defined segmentations. Furthermore, the model's training data may lack sufficient examples of certain pathological conditions, such as cardiomyopathy or spongy myocardium. This scarcity of cases can reduce the model's ability to generalize to these complex conditions, affecting its reliability in clinical settings.

Therefore, while the approach is robust under ideal conditions, its accuracy depends largely on the quality of the input data. Special care is needed when working with low-quality images or uncommon pathologies, as these can lead to decreased accuracy and make the model less reliable in critical diagnostic scenarios.

## 5. Conclusions

This study presented a novel approach to cardiac MRI segmentation and classification, significantly improving accuracy using a multi-stage process combining U-Net and ResNet models to enhance the segmentation of heart structures. Gaussian smoothing is applied to refine the contours and minimize artifacts. The classification process leverages a cascade of DL classifiers to distinguish between heart conditions such as hypertrophic cardiomyopathy, myocardial infarction, and dilated cardiomyopathy.

The performance of the methods was evaluated using the Dice coefficient for segmentation accuracy and several classification metrics. The proposed approach demonstrated significant improvements in segmentation accuracy, achieving a Dice coefficient of 0.974 for the LV and 0.947 for the RV. Classification of heart conditions also showed high results, achieving an accuracy of 96% for LV pathologies, 100% for hypertrophic cardiomyopathy, and 90% for differentiating myocardial infarction from dilated cardiomyopathy. Despite these promising results, the method has limitations, particularly when processing low-quality images or dealing with complex pathologies, where segmentation accuracy may decrease.

Future work will focus on developing new techniques for interpreting the results, aiming to make the method more applicable and reliable in clinical settings.